**Title:** Nematic colloidal micro-robots as physically intelligent systems

**Authors:** Tianyi Yao[1], Žiga Kos[2,3], Yimin Luo[4], Francesca Serra[5], Edward B. Steager[6], Miha Ravnik[2,7], Kathleen J. Stebe[1,*]

**Affiliations:** [1]Chemical and Biomolecular Engineering, University of Pennsylvania, Philadelphia, PA 19104, USA; [2]Faculty of Mathematics and Physics, University of Ljubljana, Jadranska 19, 1000 Ljubljana, Slovenia; [3]Department of Mathematics, Massachusetts Institute of Technology, Cambridge, MA 02139, USA; [4]Department of Chemical Engineering, University of California, Santa Barbara, CA 93106, USA; [5]Department of Physics and Astronomy, Johns Hopkins University, Baltimore, MD 21218, USA; [6]Mechanical Engineering and Applied Mechanics, University of Pennsylvania, Philadelphia, PA 19104, USA and [7]Condensed Matter Physics Department, J. Stefan Institute, Jamova 39, 1000 Ljubljana, Slovenia
*Corresponding Authors Emails: kstebe@seas.upenn.edu

**Abstract:** Physically intelligent micro-robotic systems exploit information embedded in micro-robots, their colloidal cargo, and their milieu to interact, assemble and form functional structures. Nonlinear anisotropic fluids like nematic liquid crystals (NLCs) provide untapped opportunities to embed interactions via their topological defects, complex elastic responses, and their ability to dramatically restructure in dynamic settings. Here we design and fabricate a 4-armed ferromagnetic micro-robot to embed and dynamically reconfigure information in the nematic director field, generating a suite of physical interactions for cargo manipulation. The micro-robot shape and surface chemistry are designed to generate a nemato-elastic energy landscape in the domain that defines multiple modes of emergent, bottom-up interactions with passive colloids. Micro-robot rotation expands the ability to sculpt interactions; the energy landscape around a rotating micro-robot is dynamically reconfigured by complex far-from-equilibrium dynamics of the micro-robot's companion topological defect. These defect dynamics allow transient information to be programmed into the domain and exploited for top-down cargo manipulation. We demonstrate robust micro-robotic manipulation strategies that exploit these diverse modes of nemato-elastic interaction to achieve cargo docking, transport, release, and assembly of complex reconfigurable structures at multi-stable sites. Such structures are of great interest to future developments of LC-based advanced optical device and micro-manufacturing in anisotropic environments.

**INTRODUCTION**

Untethered mobile micro-robots are the focus of intensive research with diverse strategies for actuation, mobility and interaction (*1–6*). Given their scale, the development of these systems has exploited and inspired research in far-from-equilibrium colloidal systems. Micro-robot mobility is achieved by various mechanisms including self-propulsion and actuation under external fields, intersecting with the field of active colloids (*7–11*). The physical dimensions of micro-robots make it challenging to integrate computational elements that imbue them with computational intelligence. Thus, micro-robots typically exploit physical intelligence (*12*) to perform essential tasks including colloidal-scale cargo capture, transport and delivery. Physical intelligence refers to diverse interactions between the micro-robot, its cargo, or domain boundaries that can be harnessed to perform useful work, often drawing on concepts at the forefront of directed colloid manipulation and assembly. For example, micro-robot motion is exploited to generate hydrodynamic interactions that dictate cargo displacement (*13*, *14*), and external electromagnetic fields are applied to generate and control colloid-micro-robot interaction (*15–17*). Such interactions can be tailored by design of micro-robot and cargo shape, material properties and those of the domain boundaries (*17*). While many studies in this arena are motivated by potential biomedical applications (*2*, *3*, *6*), there are important untapped opportunities for micro-robotics in technologically relevant environments to generate reconfigurable structures for functional metamaterials ranging from advanced optical devices (*18–21*) to energy harvesting materials (*22*, *23*).

While micro-robots are typically studied in isotropic fluids, highly anisotropic domains provide important additional degrees of freedom for designing the interactions between a micro-robot and its cargo. For example, curvature fields at fluid interfaces have been designed to direct colloid motion by capillarity (*24*); such interactions have recently been exploited for micro-robotic assembly and cargo manipulation (*25–27*). Nematic liquid crystals (NLC) are anisotropic fluids in which physical information can be embedded via the organization of nematogens and the presence of topological defects to generate emergent interaction among microscale objects in the domain (*28–36*). The introduction of micro-robot dynamics dramatically expands the opportunity to sculpt such interactions. For example, micro-robot shape and anchoring conditions can mold nematogen orientation and dictate the formation of topological defects; the ability to reposition the micro-robot allows this information to be embedded at arbitrary sites in the domain. Furthermore, in far-from-equilibrium systems, the energy landscape around micro-structures can be dynamically reconfigured to generate dynamic defect structures (*37–43*) by an interplay of the elasticity and external fields in these highly non-linear fluids. Such reconfigurable energy landscapes provide exciting opportunities for exploitation in untethered micro-robotic systems.

In this paper, we design and fabricate a 4-armed ferromagnetic micro-robot which can be actuated using an external magnetic field (Fig. 1A). The micro-robot's shape and surface chemistry are designed to embed a nemato-elastic energy landscape that generates complex force fields (Fig. 1B) on passive colloids. These emergent interactions drive the colloids along well-defined paths toward distinct sites for assembly with interactions strengths of magnitude $\sim 10^5 k_B T$. Micro-robot rotation by an external magnetic field generates far-from-equilibrium dynamics of the micro-robot's companion topological defect (Fig. 1, C-E) which provide additional opportunities for micro-robot mobility and passive cargo manipulation and assembly. In this research, we introduce and characterize these aspects of NLC micro-robotics as forms of physical intelligence and exploit them in top-down and bottom-up assembly of robust reconfigurable micro-structures.

## RESULTS
### Micro-robot design and its static defect configurations

We fabricate a ferromagnetic 4-armed micro-robot using standard lithographic methods followed by PVD sputtering of a layer of Ni ($\sim 20$ nm and $\sim 200$ nm for thin and thick coating, respectively) and subsequent treatment with dimethyloctadecyl [3-(trimethoxysilyl)propyl] (DMOAP). The resulting micro-robot has homeotropic (perpendicular) anchoring on its Ni-coated top and side surfaces, and degenerate planar anchoring on its bottom face. When placed in a uniform planar cell filled with the nematic liquid crystal 4-cyano-4'-pentylbiphenyl (5CB) (Fig. 1A), the micro-robot molds the local director field, a headless vector **n**, which describes the orientational order of the nematogens. The micro-robot's arms and wells have curvatures designed to generate gentle distortions in the domain to promote lock-and-key assembly of passive colloids (Fig. 1, A and B). Given the micro-robot's complex anchoring and sharp edges, its defect configurations differ significantly from their well-known counterparts on smooth particles with uniform anchoring.

Once placed in the planar cell, two different defect structures emerge depending on the gap thickness between the two plates. For highly confined systems where the ratio of cell thickness to micro-robot thickness $h/H$ is $\sim 1.2$, the system assumes a metastable defect configuration with a quadrupolar symmetry, with two defects at the tip of the two arms aligned perpendicularly to the far field director (Fig. S1 in SI). For less confined systems ($h/H \sim 2$), a stable configuration emerges with dipolar symmetry, with a single defect visible at the tip of one of the arms aligned parallel to the far field director as shown in Figs. 2A and 2B. Numerical simulation (Fig. 1A and Fig. 2C) reveals that this defect is a disclination loop that is anchored on two locations on the micro-robot's degenerate planar face and extends along the micro-robot's side toward its homeotropic face. The stability and shape of the dipolar structure in simulations depend on the details of the surface anchoring (Fig. S2 in SI). In experiments, the quadrupolar structure irreversibly transforms to the dipolar structure under external perturbation. In addition to these defects, simulation reveals zones of diminished order along the sharp edges of the micro-robot.

To study these micro-robots in interaction with passive colloids under weak confinement, DMOAP-treated silica colloids ($2a = 25 \mu m$) with homeotropic anchoring are suspended with the micro-robot in 5CB; this suspension is introduced into the planar cell in the isotropic state, and subsequently quenched into the nematic state by cooling below the isotropic-nematic transition temperature ($T_{IN} \sim 35.2°C$). We focus on the weakly confined case in which both the micro-robot and the colloid carry dipolar defects. Under the action of an external magnetic field (details in the Methods section), the micro-robot can translate or rotate with complex defect dynamics that are harnessed to interact with passive colloids.

**Directed assembly of colloids in nemato-elastic force field**

The micro-robot embeds a complex energy landscape in the surrounding NLC that generates emergent interactions that drive colloids along distinct paths. The path followed by a particular colloid depends on the colloid's polarity, its initial position, and the pose of the micro-robot. We enumerate the interactions that occur for a micro-robot at a fixed position with a defect on its left arm in interaction with colloids with either polarity in Figure 3. Five types of attractive interactions were observed: (i) dipole-chaining, in which the colloid chains with the dipolar loop adjacent to the robot with its companion defect pointing outward, oriented along the far field director (Fig. 3A), (ii) a rarely-observed zig-zag dipole configuration, in which the colloid assembles with its defect oriented facing the defect of the micro-robot assemble into an anti-parallel structure (Fig. 3B), (iii) a dipole-on-hill configuration, in which the colloid docks on the curved tip of the robot arm with its companion defect pointing toward the robot (Fig. 3C), and (iv) a dipole-in-well configuration, in which the colloid docks in a well between two arms of the micro-robot with its companion defect pointing outwards, oriented along the far field director (Fig. 3D), and, finally, (v) a hybrid configuration where the colloid partially docks in a well with its companion defect tilted toward the nearest arm (Fig. 3E). The dipole-chaining and zig-zag configurations are reminiscent of dipole-dipole interactions of uniform colloids with homeotropic anchoring (*44*), although the details of the defect configurations on the micro-robot differ. Note that the zig-zag dipole configuration relies on the micro-robot and colloid being in close proximity. The other three cases are a recapitulation of lock-and-key interactions in which colloids interact with gentle distortion fields seeded by the curved boundaries, as shown in the insets to Figs. 3 C-E.

Far from the micro-robot, the hedgehog companion defects on the colloids align with the far-field director with either rightward-facing or leftward-facing companion defects aligned with the far field director; this orientation defines the colloids' polarity, as shown in the insets to Fig. 3F (i) and Fig. 3G (i), respectively. Colloid-micro-robot interactions depend on this polarization and the colloid's initial position with respect to the micro-robot. The trajectories of all colloids that assembled on the micro-robot are superposed on the micro-robot frame for colloids with rightward-facing defects in Fig. 3F (i) and leftward facing defects in Fig. 3G (i). In these figures, trajectories are color-coded in terms of their final mode of assembly. For example, the green curves represent the colloids' trajectories as they approach the micro-robot and docks in the dipole-in-well configuration, while the blue curves represent the hybrid configuration in which the colloids are first repelled away from the micro-robot and then are attracted to the final equilibrium positions, resulting in complex, curved trajectories. All five assembly configurations and their trajectories are predicted by numerical analysis (Figs. 3F(ii) and 3G(ii)). We consider the free energy of point dipoles aligned along the local director field, $F_d = 4\pi K \mathbf{p} \cdot (\nabla \cdot \mathbf{n})$, where $K$ is the elastic constant and $\mathbf{p}$ is the dipolar orientation (*29*). Predicted colloid trajectories corresponding to the negative gradient in free energy are calculated for colloids with right-facing defects (Fig. 3F(i)) and left-facing defects (Fig. 3G (i)) in the far-field. Only the director gradients in the horizontal plane are considered in the analysis to avoid any influence of the director field ansatz on the sidewall of the micro-robot. These distinct modes of assembly depend strongly on initial relative positions and polarizations of the colloid and differ in range and strength, allowing for selective directed assembly.

To better understand these interactions, we track trajectories and compare the normalized separation distance to their equilibrium position $d_e / a$ as a function of time $t_c - t$, where $t_c$ is the time when the colloid reaches its equilibrium position as shown in insets to Figs. 3H-3K. The dipole-chaining and hybrid configurations have longer-ranged interaction, with attraction observed for distances as large as $6a$ from the colloids' equilibrium positions, while the dipole-on-hill and dipole-in-well configurations have attractions up to separation distances of roughly $2a$. Since colloids move with negligible inertia, i.e. with Reynolds number $Re = \dfrac{\rho u a}{\mu} \sim 10^{-7} - 10^{-6}$, where $\rho$ is the density of 5CB, $u$ is the speed of the colloid and $\mu$ is the average viscosity of 5CB, inertia can be neglected and the energy of interaction $U$ between a colloid and the micro-robot can be inferred from the energy dissipated along the colloids' trajectories. Estimates for $U$ (shown in Figs. 3 H-K) are calculated for each colloid-micro-robot configuration by integrating the viscous drag forces along the trajectories of the colloids $U = \int_{s_0}^{s_t} F_{drag}(s)ds = C_D 6\pi\mu a \int_{s_0}^{s_t} u(s)ds$, where $s_0$ is the reference point, $s_t$ is an arbitrary point along the trajectory and $C_D$ is the effective drag coefficient for this confined setting (*31*). The plots for $U$ are truncated at $d_e = 0.25a$ (dashed lines in Figs. 3 H-K) to exclude the region very near contact as near-field hydrodynamic interactions between the micro-robot and the colloid become more dominant. The stronger interactions are of magnitude $\sim 10^5 k_b T$, including the dipole-chaining, dipole-on-hill, and hybrid configurations, all of which involve the colloid's defect in their final assembled configuration. However, the relatively weak dipole-in-well configuration, of magnitude $\sim 10^4 k_b T$, relies only on matching regions of bend and splay distortion around the colloid as it docks in the well.

The slopes of the energy of interaction $U$ as the colloids approach their equilibrium positions reveal important features of the different docking configurations. While the parabolic shape of the $U$ plot for the dipole-in-well configuration suggests an elastic force on the colloid is on the order of a few $pN$ with elastic constant in the order of $10^{-7} N/m$, the more linear relationships between $U$ and separation distance $d$ for the other assembled configurations suggest that these structures require a force in excess of a yield force on the order of $10^1 - 10^2 \ pN$ to separate the colloid from the micro-robot. The distinct strengths and characteristic behaviors of the different docking configurations have important implications in terms of the ability of the micro-robot to retain and carry colloidal cargo as it moves about in the domain. For example, the elastic nature of the dipole-in-well configuration implies that the colloidal cargo will be displaced by viscous drag as the micro-robot moves, causing it to be lost from the binding pocket and left in the bulk (Figs. S3 a and c). In contrast, the colloid assembled on the micro-robot via the strongly attractive modes like the dipole-chaining configuration move together with the micro-robot (Figs. S3 b and c) as long as the hydrodynamic drag does not exceed the yield force. While such configurations allow the micro-robot to capture, retain, and transport colloidal cargo, their very strength may impede cargo release. This motivates our exploration of far-from-equilibrium defect structures.

**Far-from-equilibrium defect dynamics**

Under a rotating magnetic field, a far-from-equilibrium defect emerges whose dynamics are influenced by the micro-robot's hybrid anchoring conditions, complex shape, and sharp edges. Under slow rotation, the dipolar loop, situated at rest on a micro-robot arm aligned with the far field director, extends as it is placed in an antagonistic orientation with respect to the anchoring imposed on the planar bounding surfaces of the cell. At higher rotation rates, backflow becomes significant in addition to this geometric frustration, as the director field and flow field become coupled to leading order. Upon rotation by $\frac{\pi}{2}$ radians, the defect becomes unstable and hops to the arm that has become aligned with the far field as shown in Fig. 4A. As the defect hops, the dipolar structure interacts with a portion of the disclination loop beneath the micro-robot. When rotation ceases, the defect retracts to the dipolar configuration on the arm aligned with the director over time scales characterized by the relaxation dynamics of the system, given by $\tau = \frac{L^2 \gamma_1}{K}$, where $L$ is the characteristic length of the micro-robot, $\gamma_1$ is the rotational viscosity of NLCs. Under continuous slow rotation, the defect 'travels' via periodic extension, interaction, hopping and contraction in the opposite sense of the micro-robot's rotation. For the example, as shown in Fig. 4A, where the Erickson number $Er = 1.55$. Similar behavior has been observed for $Er$ as low as 0.06. Numerical simulations suggest distinct modes of defect hopping. Simulations of the rotating micro-robot with a dipolar loop (Fig. 4B) show dynamic elongation of the loop and sliding of the defect's pinning point on the micro-robot's edge. This sliding mechanism allows the loop to move from one arm to another. Simulations of a micro-robot with a dipolar hedgehog companion defect (Fig. S4 in SI) show another mode of hopping. In this case, the defect hops between neighboring arms of the micro-robot through the bulk phase due to strong dynamic alignment of the director field. Structurally, both pathways for defect hopping occur in regions of large director field deformations, and the motion of the defect reduces the elastic free energy of the system. Both experiment and simulation capture hopping of the defect between adjacent arms on the rotating micro-robot, as the system attempts to preserve the overall dipolar orientation with respect to the far-field. Under continuous rotation at angular velocity $\omega$ at rates that challenge the natural relaxation dynamics, the defects form smeared-out dynamic structures with periodic rearrangements that depend on the Ericksen number $Er = \omega\tau$. Under high $Er \sim 18$, for example, the defect becomes significantly elongated and remains 'smeared out' while hopping along the structure every $\frac{\pi}{2}$ radians determined by the geometric symmetry of the 4-armed micro-robot, lagging the alignment of the arm tips without relaxing back to the equilibrium structure. The extent of defect elongation is positively related to $Er$; larger $Er$ leads to greater defect elongation around the micro-robot. These defect dynamics play a very important role when interacting with colloidal cargo.

**Cargo juggling and release**

The emergence of far-from-equilibrium defects during micro-robot rotation are evidence of non-linear restructuring of the director field and provide an important means to manipulate colloidal cargo. Figure 5A shows a colloid assembled in the upper left well in the hybrid configuration at $t = 0s$. As the micro-robot rotates, the assembled colloid moves around the micro-robot influenced by steric hindrance and hydrodynamics and encounters the micro-robot's dynamic defect at $t = 25s$ after $\frac{3}{2}\pi$ rotation. The elongated disclination loop of the micro-robot and the hedgehog defect from the colloid then merge to form a shared defect that carries the colloid along with it at $t = 29s$. Upon further rotation, the shared defect separates to restore the colloid's companion hedgehog defect and the micro-robot's elongated defect ($t = 35s$). These dynamics place the colloid in a repulsive configuration on the hill of the arm adjacent to its initial docking site, with its defect pointing outward; the colloid is repelled from the micro-robot ($t = 47s$). Colloids assembled in other configurations can also be released by rotation via similar defect dynamics.

In another example, two identical colloids docked on the micro-robot can be juggled, rearranged, and restructured by far-from-equilibrium defect dynamics. The micro-robot with assembled colloids, shown in Fig. 5B at $t = 0s$, is initially at rest, with colloid 1 (shown in red) docked in the bottom left well in the hybrid configuration and colloid 2 (shown in yellow) docked in the upper right well in the dipole-in-well configuration. As the micro-robot rotates in the clockwise direction, its defect lags behind the arm, elongates and merges with colloid 1's companion defect to form a shared structure ($t = 19s$). This merged, shared defect carries colloid 1, rearranging its orientation and position on the micro-robot ($t = 46s$). Thereafter, the merged defect further elongates along the sharp edges of the micro-robot and encounters the companion defect of colloid 2 ($t = 66s$), forming a merged defect that is now shared with both colloids. This larger loop also rotates and re-positions colloid 2. Finally, as shown in Fig. 5B at $t = 77s$, the merged defect becomes unstable and contracts; the micro-robot recovers its original defect structure, and the colloids' companion hedgehog defects are restored. However, the changes in the positions and orientations of the colloids alter their ensuing interactions with the micro-robot. Colloid 1 eventually stably docks in the dipole-in-well configuration on the upper right. Colloid 2, however, placed in an antagonistic orientation at the tip of the arm with its hedgehog defect pointing outward, moves away from the micro-robot. Thus, colloid 1 is retained and colloid 2 is repelled from $t = 77s$ to $t = 135s$.

**Micro-robotic directed assembly of colloidal structures**

Having demonstrated the ability to assemble, transport and release passive cargo using our micro-robot, we further exploit the micro-robot for assembly of colloids and build structures by releasing these colloids near attractive sites on wavy walls (*31*, *34*, *35*). For example, a colloid in a dipole chaining configuration (Fig. 6A) was carried as cargo by a micro-robot. This cargo was released near an attractive well on a wavy micro-structure via rotational defect dynamics like those described above, including defect elongation, merger, separation, and recovery. Once detached from the micro-robot, the colloid migrates into the attractive well (Fig. 6A (iii)) and the micro-robot is driven away to retrieve a different colloidal building block in the domain. This process, which combines top-down direction by the micro-robot motion and bottom-up assembly via the emergent interactions between the colloid, the micro-robot and the wall, can be repeated and multi-element systems can be built. Depending on the design of the attractive sites, various colloidal structures can be constructed by the sequential addition of colloidal building blocks to prescribed sites. For example, using a bounding wavy wall as a construction site, as shown in Fig. 6A, multiple structures were constructed including a 1D colloidal lattice (Fig. 6B), a chain of 7 colloids (Fig. 6C) and a more complicated anisotropic structure (Fig. 6D). Our approach is material-independent as the nemato-elastic energy field is only dictated by the surface anchoring of the micro-robot and passive cargo which is defined by the initial surface treatment process and can be easily controlled. Thus, such micro-robotic assembly approach can be applied to functional building blocks made of differing materials for reconfigurable devices.

**Trajectory planning of micro-robot and fully autonomous cargo manipulation**

The generation of strong magnetic field gradients on the micro- and smaller scales remains challenging and would hamper efforts to scale down this system to manipulate colloids of smaller radius. To address this issue, we exploit a defect-propelled swimming modality of nematic colloids using a purely rotating external field to actuate the micro-robot toward fully autonomous cargo manipulation (*45*). Upon rotation, the companion defect of the micro-robot undergoes periodic rearrangement in which the defect depins from the micro-robot's sharp edge and sweeps across the surface of the micro-robot; this occurs even as the defect hops between the micro-robots' arms as shown in Fig. 7A. This defect sweeping motion acts as a swim stroke that drives micro-robot swimming; via this effect, micro-robot rotation generates unidirectional translation. For example, the micro-robot with its top arm initially positioned in contact with the dashed line in Fig. 7A traveled a distance of $30.3 \mu m$ in one period of rotation ($T = 160s$). Translational speed and direction are controlled by the rate and sense of rotation, as described in a detailed study of a rotating magnetic disk with similar hybrid anchoring in a separate study from our group (*45*). We use this modality to actuate and control the micro-robot trajectories as shown in Figs. 7B and 7C. Mirror symmetric changes in micro-robot trajectory with respect to the axis perpendicular to the far field director were achieved by reversing the micro-robot's sense of rotation as shown in Fig. 7B. The direction of the external field of $T = 12s$ was reversed twice at the locations indicated by the red dashed lines which led to a N-shaped trajectory of the micro-robot. More complex changes of direction are also possible. The direction of micro-robot translation depends on the rate of rotation, affording an additional degree of control. As shown in Fig. 7C, by changing the period of rotation from $4s$ to $8s$ and finally to $16s$ at the locations indicated by the red dashed lines, the micro-robot moves along a curved trajectory. This ability to steer the micro-robot relies on the rotation-rate dependent defect elongation which enhances broken symmetry in the system. With the ability to steer and make sharp turns while translating purely by tuning the rotation rate and direction, the micro-robot can be exploited for fully autonomous micro-robotic cargo manipulation.

We demonstrate fully autonomous micro-robotic cargo manipulation using our 4-armed micro-robot under a programmable rotating magnetic field as shown in Fig. 8. The complete process was divided into four stages: (i) locomotion and approach, (ii) directed assembly, (iii) transport and (iv) release. Firstly, the micro-robot was driven towards a colloid, following an almost linear path of $\sim 137 \mu m$ with an average speed of $1.71 \mu m/s$ under a clockwise rotating external field of $T = 4s$ (Fig. 8(i)). Upon cessation of rotation, the micro-robot recovers its static dipolar defect and the colloid is attracted and migrates a distance $\sim 5.6a$ to dock in the dipole chaining configuration (Fig. 8(ii)). Note that the time required for this docking process depends on the initial separation distance between micro-robot and colloid which determines the strength of the elastic interaction; this time can be greatly reduced if the micro-robot is placed closer to the colloid. Here, we parked the micro-robot relatively far from the colloid in order to demonstrate the range of this interaction. Once assembled, the pair was rotated counterclockwise under the same period of $T = 4s$. During this process, the micro-robot followed a linear path while the colloid follows a helical trajectory and travels an effective distance $\sim 16.4a$; the retention of the colloid is influenced by a complex interplay of defect-defect interaction and hydrodynamics (Fig. 8(iii)). Finally, upon reducing the period of rotation to $T = 20s$, the extent of defect elongation is reduced, weakening the attractive interactions between the colloid and micro-robot's defects. The colloid is released from the micro-robot (Fig. 8(iv)), completing this fully autonomous cargo manipulation process.

## DISCUSSION

We have introduced the concept of driven micro-robots in NLC as physically intelligent systems imbued with the capability to sense, attract and assemble colloidal building blocks via material agnostic nemato-elastic interactions and to dynamically restructure their environment. This untethered micro-robotic platform in NLC can generate complex colloidal reconfigurable structures via a combination of top-down and bottom-up assemblies. The motion of micro-robots in NLC is strongly coupled to the highly anisotropic nematic organization, and vice versa, providing opportunity to dramatically reconfigure the elastic energy landscape and to write transient director fields into the domain for potential micro-robotic applications.

Here we have described micro-robots with shapes and surface chemistry designed to embed elastic energy landscapes and generate distinct emergent interactions with colloidal cargo. Furthermore, the micro-robot's rotational motion can deform its companion topological defect to generate rich non-equilibrium defect dynamics. We have exploited such dynamics as virtual functional structures that generate modalities of motion and interaction to enable reconfigurable assembly of passive building blocks with remarkable degrees of freedom. Finally, we have demonstrated a fully autonomous cycle of cargo manipulation using a swimming modality enabled by the dynamic defect, which propels micro-robot translation. This ability to generate dynamic force fields, dynamically restructure the topological defects and exploit them as functional structures for colloidal assembly greatly expands the opportunities for assembly of reconfigurable functional systems. We envision applications ranging from functional metasurfaces and devices to manage electromagnetic, including thermal, fields. Our approach, which exploits the nematic liquid crystals' anisotropic response to generate micro-robot-colloidal cargo interactions differs from existing approaches for reconfigurable devices that exploit nematic liquid crystal's optical birefringence. Should our approaches gain traction, the opportunity for impact is vast, as society has made tremendous investment in the grooming of liquid crystalline responses, for example, in the over $160B/year thin-film transistor (TFT) liquid crystal display industry (*46*).

The field of micro-robotics has spurred advancements in far-from-equilibrium soft matter colloidal physics. In this research, the highly nonlinear dynamic response of nematic liquid crystals revealed by the micro-robot's motion has generated open fundamental questions that are worthy of detailed study. For example, the micro-robot has hybrid anchoring and rough sharp edges whose impact on dynamic defect pinning/depinning and defect elongation thresholds remain to be elucidated. The elongated defect undergoes multiple complex rearrangements including the swim stroke and defect hopping instabilities whose dependence on micro-robot properties and rotational dynamics warrant further study. In far-from-equilibrium micro-robot/cargo interactions, transient defect-defect interactions including defect sharing, merger and separation play central roles in cargo fate. The settings in which defects remain distinct, merge, or re-separate in these highly non-linear regimes are not known, and the physics that regulates these transitions remains unexplored. Greater fundamental understanding of such far-from-equilibrium behaviors would further develop NLC micro-robot physically intelligent interactions and would advance such system's potential for tether-free micro-robotic cargo manipulation in technologically relevant settings. Finally, our results may spur research in active nematic systems such as microtubules whose nematogens consume chemical energy and dynamically reconfigure (*47*, *48*). The emergent behaviors harnessed in our system originate from the nematic fluid's anisotropy; related effects should also emerge in active nematic systems, with the additional potential of harnessing the activity of these systems to drive micro-robot motion and interaction.

## MATERIALS AND METHODS

### Fabrication of micro-robots and assembly of planar NLC cell

Micro-robots with critical dimensions shown in Fig. 1A were fabricated out of SU-8 photoresist (Kayaku Advanced Materials, Inc.) following lithographic processes on a supporting wafer. Thereafter, a layer of nickel was sputtered onto the surface using a Lesker PVD75 DC/RF Sputterer to make the colloids ferromagnetic. Subsequently, treatment with $3wt\%$ solution of N-dimethyl-n-octadecyl-3-aminopropyl-trimethoxysilyl chloride (DMOAP, Sigma-Aldrich) imposed homeotropic anchoring condition on the disk's Ni coated surfaces. Silica spherical colloids of $2a = 25\mu m$ (Spherotech Inc.) were also treated with DMOAP, washed and dried before adding into 4-cyano-4'-pentylbiphenyl

(5CB, Kingston Chemicals). Finally, the micro-robots were released from the wafer and dispersed in a suspension of passive colloids in 5CB. Glass slides were spin-coated with polyimide (PI-2555, HD Microsystems) and rubbed with a velvet cloth along the desired direction to impose uniform planar anchoring. Two glass slides with uniform planar anchoring were assembled in an antiparallel fashion and glued together using a UV sensitive epoxy with two layers of $15\mu m$ plastic spacers in between. The resulting thickness of the cell was $\sim 50\mu m$. Finally, a mixture suspension of 5CB with micro-robots and passive colloids was introduced into the cell from the side by capillarity in the isotropic state of 5CB before quenching down to the nematic state. Depending on the thickness of the nickel layer, the coated micro-robot could either appear transparent (nickel layer $\sim 20nm$) or black (nickel layer $\sim 200nm$). While the transparent micro-robot allowed us to visualize the sweeping motion of the disclination line, micro-robots with thicker coating possess stronger magnetic moments, enabling faster rates of rotation and translation under external magnetic fields.

**Application of external magnetic fields**

Controlled rotations of the micro-robots were achieved by placing the assembled NLC cell in a rotating magnetic field generated by a custom-built magnetic control system. The system consists of two orthogonal pairs of electromagnetic coils (APW Company) mounted on an aluminum supporting structure arranged around the workspace. Visual feedback is provided by a CCD camera (Point Grey Grasshopper3 Monochrome) mounted on a Zeiss inverted microscope (ZEISS Axio Vert.A1). Each coil pair was powered independently using a programmable power supply (XG 850W, Sorensen) whose outputs were controlled by a Python algorithm written in-house. Sinusoidal time-dependent voltages are applied on each pair and the waveforms are separated by a $\frac{\pi}{2}$ phase lag to achieve a rotating field. The field gradient was applied by using a rectangular NdFeB magnets (K&J Magnetics, Inc.) held to the end of a tweezer. The magnet was placed roughly ~0.5 cm from the cell. The amplitudes of the magnetic field applied are measured using a magnetometer and are in the order of a few $mT$, far below the magnetic Fréedericksz transition threshold to reorient the NLC molecules, but sufficiently strong to overcome the drag and move the micro-robot in arbitrary directions.

**Details on numerical modeling**

Numerical simulations of static and dynamic nematic structures were performed using a Q-tensor order parameter formulation of nematodynamics. The scalar degree of order $S$ and the director $\mathbf{n}$ are the largest eigenvalue and the corresponding eigenvector of the Q-tensor, respectively. Equilibrium configurations correspond to minima of the Landau-de Gennes free energy with volume density of

$$f_{\text{vol}} = \frac{A}{2}Q_{ij}Q_{ji} + \frac{B}{3}Q_{ij}Q_{jk}Q_{ki} + \frac{C}{4}\left(Q_{ij}Q_{ji}\right)^2 + \frac{L}{2}(\partial_k Q_{ij})(\partial_k Q_{ij}), \qquad (1)$$

where $A$, $B$, $C$ are phase parameters that dictate the degree of order in equilibrium homogeneous director field $S_{\text{eq}}$, and $L$ is the elastic constant. Additionally, Fournier-Galatola planar-degenerate surface potential describes the anchoring of nematic molecules on the bottom surface of the active micro-robot

$$f_{\text{surf}} = W\left(\tilde{Q}_{ij} - \tilde{Q}_{ij}^{\perp}\right)^2,\qquad(2)$$

where $\tilde{Q}_{ij} = Q_{ij} + \frac{S_{\text{eq}}}{2}\delta_{ij}$, $\tilde{Q}_{ij}^{\perp} = (\delta_{ik} - v_i v_k)\tilde{Q}_{kl}(\delta_{lj} - v_l v_j)$, and $\vec{v}$ is the surface normal. On the micro-robot's side wall, top surface and cell's top and bottom boundaries, the director field is fixed.

Equilibrium structures are found by using a gradient descent for the Q-tensor

$$\dot{Q}_{ij} = \Gamma H_{ij},\qquad(3)$$

where $H_{ij}$ is the molecular field $H_{ij} = -\frac{1}{2}\left(\frac{\delta F}{\delta Q_{ij}} + \frac{\delta F}{\delta Q_{ji}}\right) + \frac{1}{3}\frac{\delta F}{\delta Q_{kk}}\delta_{ij}$ and $\Gamma$ is the rotational viscosity parameter. On the planar degenerate surface, Q-tensor follows the dynamics of

$$\dot{Q}_{ij}^{\text{surf}} = \Gamma_{\text{surf}}\left[\frac{1}{2}(H_{ij}^{\text{surf}} + H_{ji}^{\text{surf}}) - \frac{1}{3}\delta_{ij}H_{kk}^{\text{surf}}\right],\qquad(4)$$

where $\Gamma_{\text{surf}}$ is the surface rotational viscosity parameter, and $H_{ij}^{\text{surf}} = -\frac{\partial f_{\text{vol}}}{\partial(\partial_k Q_{ij})}v_k - \frac{\partial f_{\text{surf}}}{\partial Q_{ij}}$ is the surface molecular field.

Simulations for the rotating micro-robot were solved in the rotating frame of the micro-robot, in which case the time derivative of the Q-tensor includes an additional term of $Q_{ik}\Omega_{kj} - \Omega_{ik}Q_{kj}$, where $\Omega_{ij}$ is the vorticity tensor of the rotating colloid corresponding to $Er \approx 6$.

Equation (3) was solved using a finite difference method on a 800x800x240 mesh. The dimensions of the micro-robot were $H = 75\Delta x$, $r_1 = 45\Delta x$, and $r_2 = 37.5\Delta x$ in accordance to Fig. 1A. Neumann boundary conditions are used in the lateral directions of the numerical simulation box. Mesh resolution is set to $\Delta x = 1.5\xi_N = 1.5\sqrt{L/(A + BS_{\text{eq}} + \frac{9}{2}CS_{\text{eq}}^2)}$, where $\xi_N$ is the nematic correlation length that sets the size of the defect cores. The following values of the model parameters are used: $B/A = 12.3$, $C/A = -10.1$, $W = 0.5L/\Delta x$, $\Gamma_{\text{surf}} = \Gamma/\Delta x$, and a timestep of $0.1(\Delta x)^2/(\Gamma L)$.

## Statistical analysis

Strengths of micro-robot-colloid interaction were determined from at least three independent experiments for each mode for statistical significance.

**Supplementary Materials**
  Text
  Fig. S1. Quadrupolar defect structure on the micro-robot.
  Fig. S2. Stable dipolar structures in numerical simulations around the micro-robot.
  Fig. S3. Cargo carrying with dipole-in-well and dipole-chaining.
  Fig. S4. Hopping of a dipole point defect during rotation of the micro-robot.

**Acknowledgments**

**Funding:**
This work was carried out in part at the Singh Center for Nanotechnology, which is supported by the NSF National Nanotechnology Coordinated Infrastructure Program under grant NNCI-2025608. Slovenian Research Agency (ARRS) under contracts P1-0099 (Ž.K. and M.R.) and N1-0124 (Ž.K.).

**Author contributions:** T. Y., Y.L., F.S and K.J.S. designed research. T.Y. and E.B.S. developed control cell; T.Y. performed experiments and analyzed experimental data. M.R. and Z.K. designed and conducted numerical simulations. T. Y. and Z. K. contributed to figure preparation. All authors contributed to the writing of the manuscript and participated in discussions of the research.

**Competing interests:** Authors declare that they have no competing interests.

**Data and materials availability:** All data needed to evaluate the conclusions in the paper are present in the paper and the Supplementary Materials. Data will be available upon request.

**Figures:**

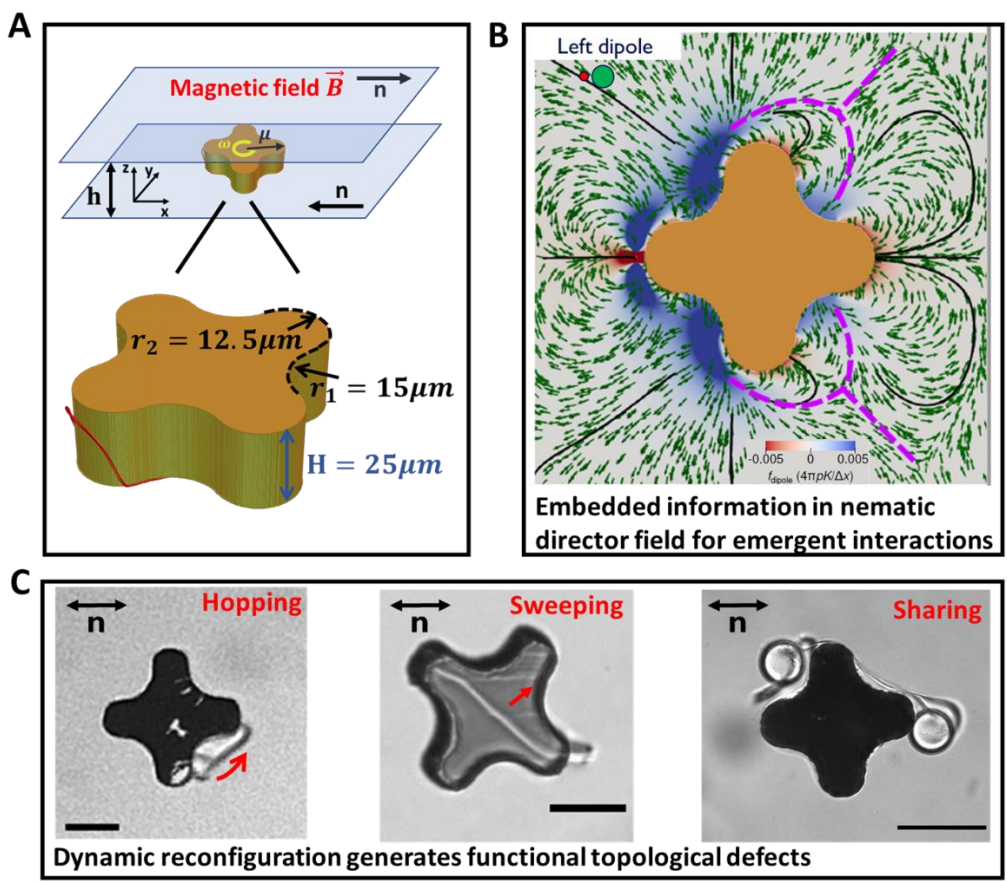

**Fig. 1. Physically intelligent magnetic micro-robot in nematic liquid crystals (NLCs).**
(**A**) Schematic of the micro-robot in a planar cell filled with NLCs. **n** denotes the rubbing direction. The thickness of the cell is denoted as $h$ and $\sim 50\mu m$ for this study. $r_1 = 15\mu m$ and $r_2 = 12.5\mu m$ denote the curvature of the micro-robot whose thickness is $H \sim 25\mu m$. (**B**) Emergent force field (green vector field) on a colloid with a left-facing companion defect near the micro-robot due to the embedded nemato-elastic energy landscape. The magenta lines indicate separatrices in the force field, indicating complex boundaries between different modes of interaction. (**C**) Far-from-equilibrium defect dynamics, including hopping, sweeping and sharing with colloidal cargo provide dynamic functionalities for micro-robot and cargo manipulation. Scale bars are $50\mu m$.

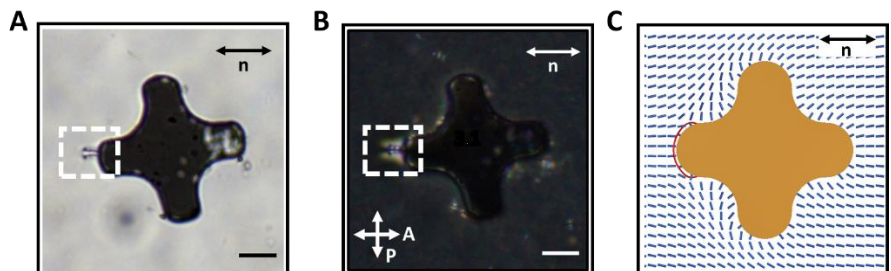

**Fig. 2. Static dipolar defect structure.** Static dipolar defect structure of the 4-armed micro-robot under bright field microscopy (**A**), crossed polarization microscopy (**B**) and in numerical simulation (**C**). The far-field director is along the horizontal direction indicated by the double headed arrows and the scale bars are $20\mu m$.

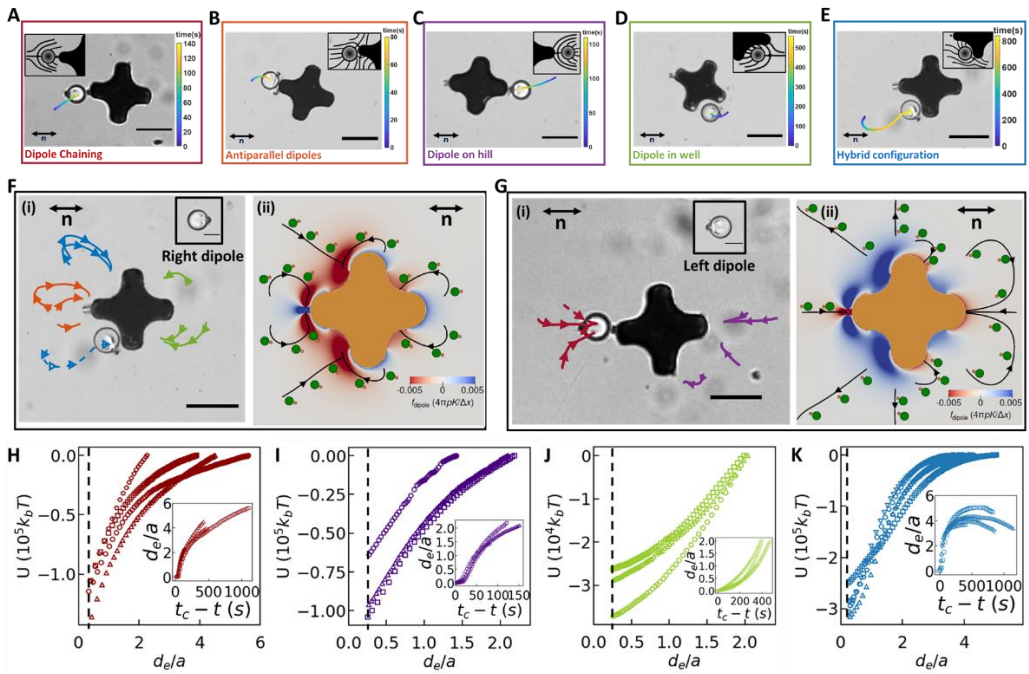

**Fig. 3. Assembly of colloids by nemato-elastic force field around the 4-armed micro-robot.** Microscopic images of different assembly configurations: dipole-chaining (**A**), zig-zag dipoles (**B**), dipole-on-hill (**C**), dipole-in-well (**D**) and hybrid (**E**) configurations around the 4-armed micro-robot. (**F**), (**G**) Summary of the experimentally observed colloid trajectories (left) and corresponding numerical predictions (right) in the micro-robot fixed frame are shown for right-facing companion hedgehog defects (F) and left-facing companion hedgehog defects (G) shown in the insets. Trajectories of different colors represent different final assembly configurations: dipole-chaining (red), zig-zag dipoles (orange), dipole-on-hill (purple), dipole-in-well (blue) and hybrid (blue) configurations. The dashed curves indicate trajectories of the colloids in the present figures. (**H**)-(**K**) Strengths and ranges of interactions of the assembly modes; here, different symbols in each plot indicate different observations. The selection among different modes, and the deviations between different trajectories are mainly caused by different initial relative positions of the micro-robot and colloid. The insets to (A)-(E) show schematics of the director fields around the colloids and the insets in (H)-(K) shows the strengths of the attractive interactions, respectively. Scale bars are $20\mu m$ for the insets in (F) and (G) and are $50\mu m$ otherwise.

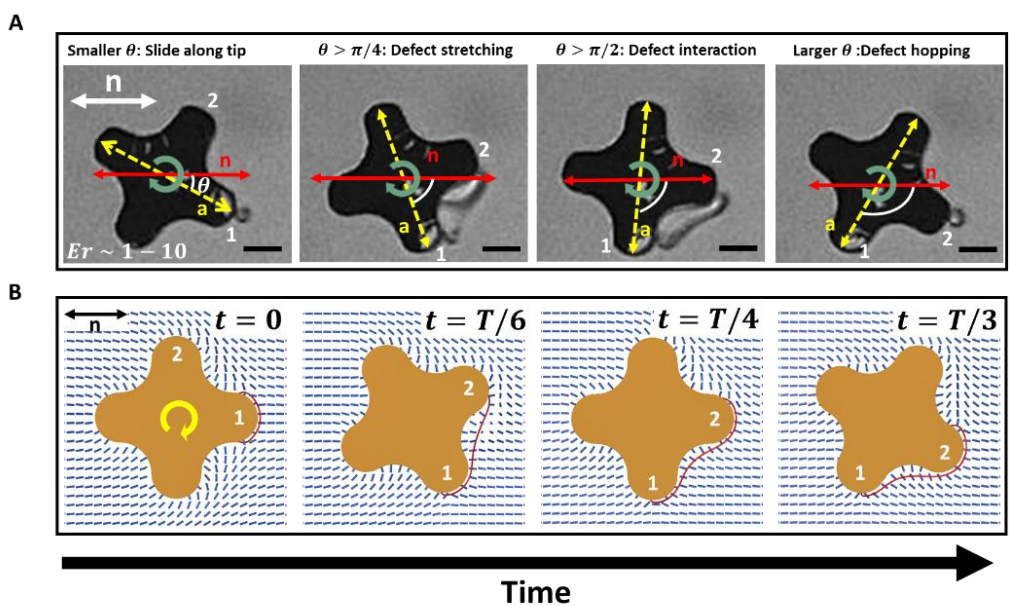

**Fig. 4. Defect hopping around the 4-armed micro-robot.** The defect exhibits a hopping instability. (**A**) Experimental time-series images of defect hopping from one arm (labeled as 1) to another arm (labeled as 2) of the 4-armed micro-robot during a $\pi/2$ clockwise rotation. $\theta$ indicates the angle between the director (red arrow) and the diagonal (dashed yellow arrow) of the micro-robot. Scale bars are $20\mu m$. (**B**) Time-series images of numerical simulation of the dipolar disclination loop defect hopping between the two arms of the 4-armed micro-robot (labeled as 1 and 2) during clockwise rotation. The red curves represent for the companion topological defect.

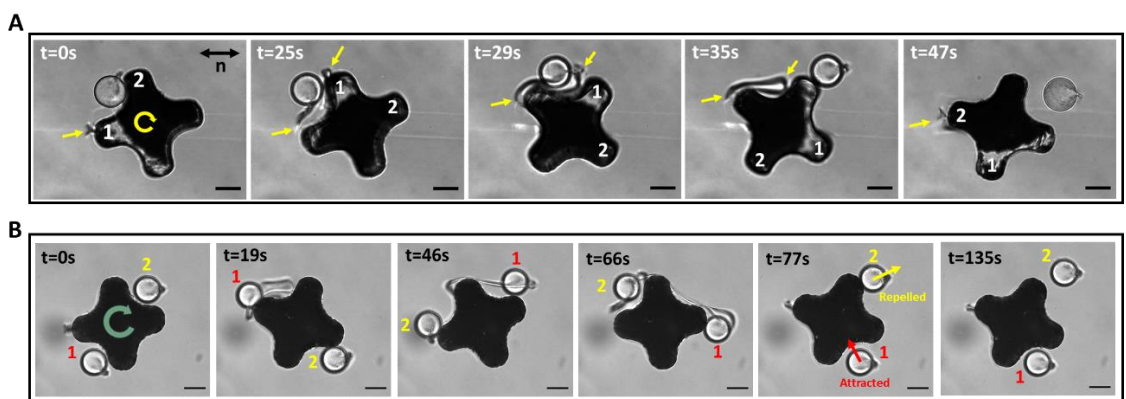

**Fig. 5. Cargo juggling and release.** (**A**) Time-stamped images of cargo release of an assembled colloid between two arms of the micro-robot (labeled as 1 and 2) during clockwise rotation via dynamic defect interaction. The yellow arrows indicate the location of the (elongated) disclination loop of the micro-robot. (**B**) Time-stamped images of cargo juggling of two colloids (labeled as 1 and 2) during clockwise rotation of the micro-robot via dynamic defect interaction. Scale bars are $20\mu m$.

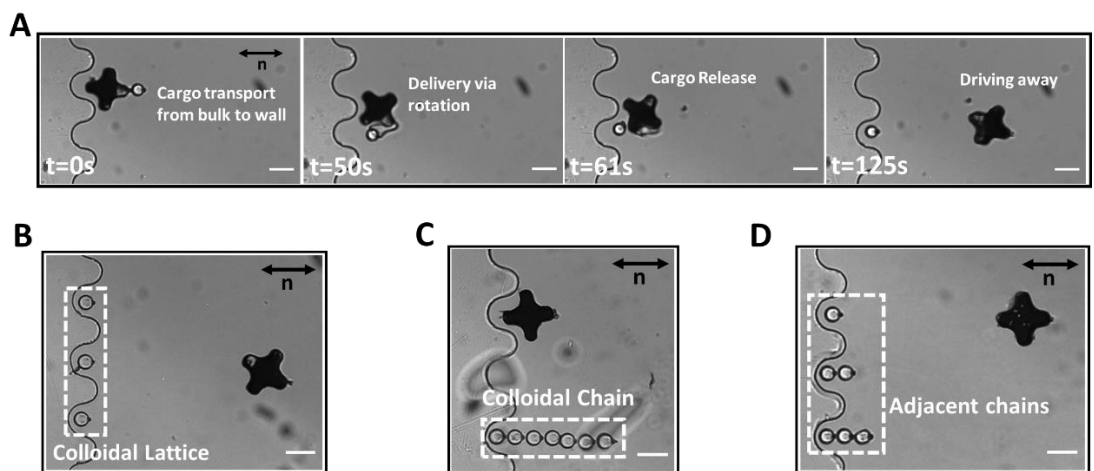

**Fig. 6. Micro-robotic directed assembly of colloidal building blocks.** (**A**) Time-stamped images of micro-robotic delivery of colloidal building block near a wavy wall exploiting top-down and bottom-up interactions. (**B**)-(**D**) Examples of colloidal structures built using our approach include a 1D colloidal lattice (B), a chain of 7 colloids (C) and an anisotropic structure resembling the signal intensity symbol on cell phones (D). Scale bars are $50 \mu m$.

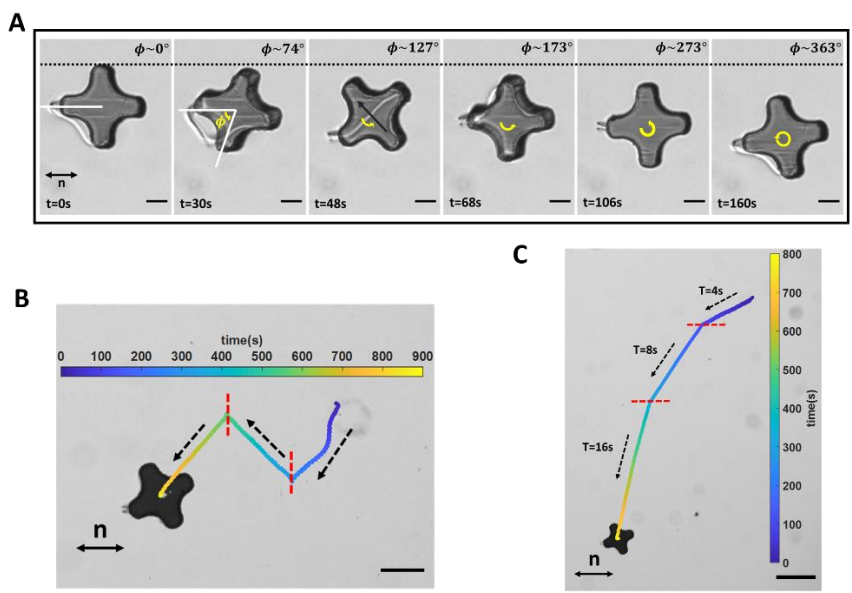

**Fig. 7. Defect-propelled micro-robot swimming and trajectory planning.** (**A**) Time-stamped images of defect-propelled swimming of the micro-robot under a clockwise rotating external field of $T = 160s$. Besides hopping, a disclination line sweeps across the surface of the micro-robot along the direction, shown by the black arrow in the frame at $t = 80s$, which propels the micro-robot. The black dashed line across all frames indicates the initial top arm tip position of the micro-robot. Trajectory planning of the micro-robot: (**B**) A N-shaped trajectory of the micro-robot achieved by reversing the sense of rotation of the external field at the locations indicated by the red dashed lines. The initial field is rotating counterclockwise with a period $T = 4s$. (**C**) A curved trajectory of the micro-robot achieved by tuning the period of the external field from $T = 4s$ to $T = 8s$ and then from $T = 8s$ to $T = 16s$ after 10 periods at the locations indicated by the red dashed lines, respectively. For both (B) and (C), the black dashed arrows indicate the direction of the translation and the colored curved indicates the trajectory of the micro-robot as a function of time. Scale bars are $50\mu m$ in (A) and (B) and is $100\mu m$ in (C).

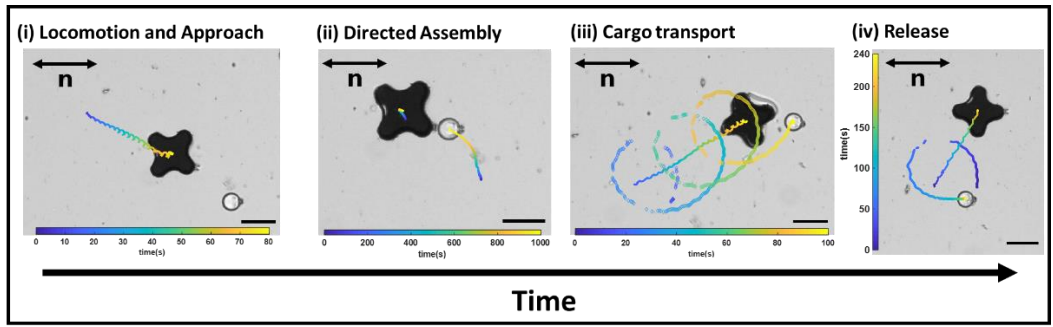

**Fig. 8. Fully autonomous cargo manipulation.** Fully autonomous cycle of cargo manipulation using our nematic micro-robot. Scale bars are $50\mu m$.

# Supplementary Information for

## Nematic colloidal micro-robots as physically intelligent systems

Tianyi Yao, Žiga Kos, Yimin Luo, Francesca Serra, Edward B. Steager, Miha Ravnik, Kathleen J. Stebe[*]

Correspondence to: kstebe@seas.upenn.edu

**This Supplementary information includes:**
    Text
    Figs. S1 to S4

The highly non-linear nature of nematic liquid crystals allows multiple (meta)stable states with distinct director fields and defect configurations to arise that depend subtly on the details of system initial conditions, degree of confinement, and anchoring conditions on the micro-robot surface and the boundaries of the domain. In Figures S1 and S2 we report static defect configurations. In Figure S3 we report simulated defect hopping for a dipolar point defect.

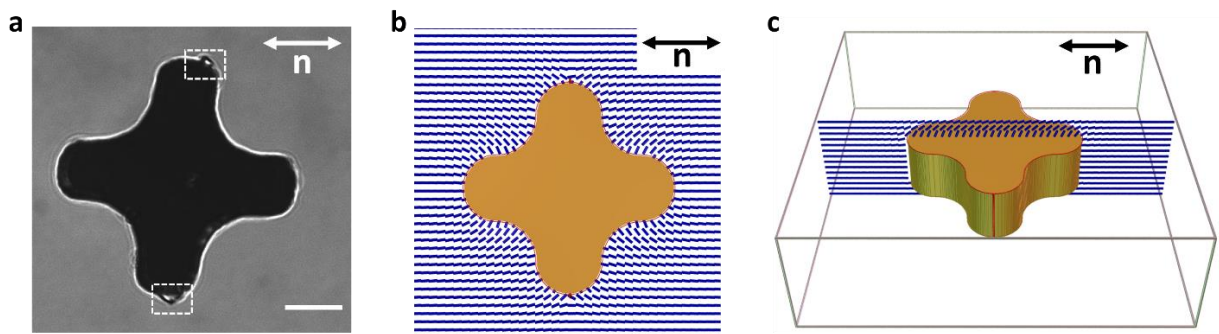

**Fig. S1. Quadrupolar defect structure on the micro-robot.**
Microscopic image (**a**) and numerical simulation (**b**) and (**c**) showing the quadrupolar defect configuration around a micro-robot. In the simulation, the micro-robot has homeotropic anchoring at the top and the side surface, and planar degenerate anchoring at the bottom surface. In the quadrupolar configuration, a defect line is pinned to the bottom and the top edge at each side of the particle. Double-headed arrows indicate the far-field director. Scale bar is $20\mu m$.

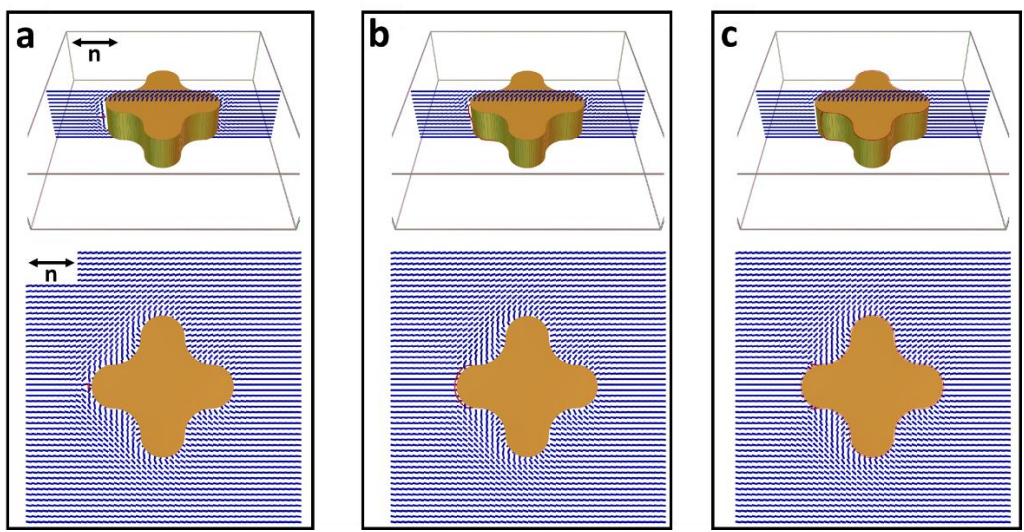

**Fig. S2. Stable dipolar structures in numerical simulations around the micro-robot.**
A splay-like director ansatz is used on the micro-robot sidewall to stabilize three distinct defect configurations and director fields, reported in (**a**)-(**c**). (a) The tilt angle out of the horizontal plane for the director goes from $-\frac{\pi}{2}$ to $\frac{\pi}{2}$ betweeen the bottom and the top surface of the micro-robot. (b) The director tilt angle is zero at the bottom half of the sidewall and goes from 0 to $\frac{\pi}{2}$ from the midplane to the top surface. (c) The director is perpendicular to the surface across the entire sidewall.

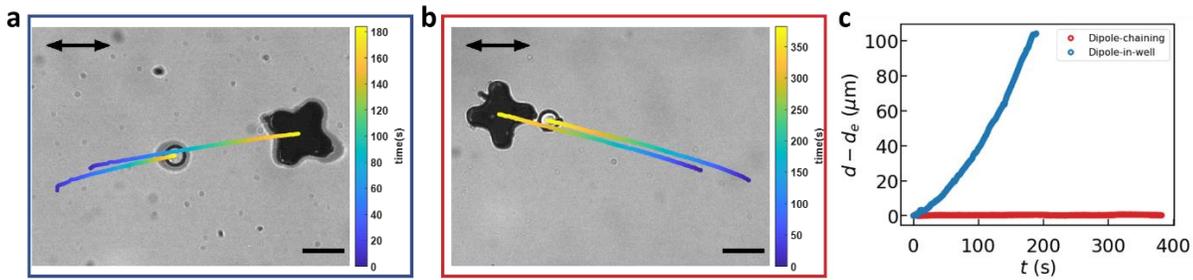

**Fig. S3. Cargo carrying with dipole-in-well and dipole-chaining.**
(**a**) Trajectories of a moving micro-robot and a colloid assembled in the dipole-in-well configuration.
(**b**) Trajectories of a moving micro-robot and a colloid assembled in the dipole-chaining configuration. Scale bars are $50\mu m$. (**c**) Calibrated center-to-center distance $d - d_e$ as a function of time indicated by the colored-curves in (a) and (b), where $d_e$ is the equilibrium center-to-center distance before the micro-robot started to translate.

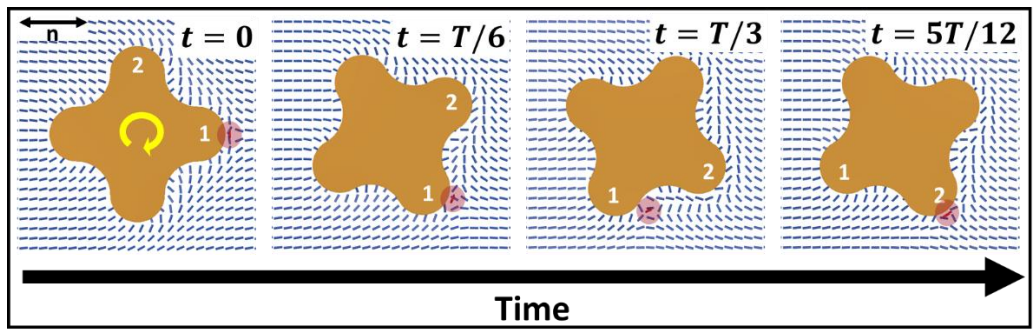

**Fig. S4. Hopping of a dipole point defect during rotation of the micro-robot.**
Two arms of the micro-robot are labelled as 1 and 2 for easier visualization of defect motion. The dipolar point defect (shadowed in dark red), obtained by applying a splay-like director ansatz across the sidewall of the micro-robot, is located on the tip of arm 1 absent micro-robot rotation. Under clockwise rotation of the micro-robot, the dipolar point defect hops from arm 1 to arm 2 by travelling through the bulk liquid crystal adjacent to the micro-robot.